\documentclass[apj]{emulateapj}

\usepackage{natbib}
\usepackage{graphics}
\usepackage{booktabs}
\usepackage{epsfig}
\newcommand{\bp}{$\beta$~Pic}
\newcommand{\si}{\emph{Spitzer}}
\newcommand{\iras}{\emph{IRAS}}
\newcommand{\lir}{$L_\mathrm{IR}/L_\star$}
\newcommand{\um}{$\mu\rm{m}$}

\slugcomment{Draft version November 28, 2007. Accepted for publication in ApJ.}

\shorttitle{Debris Disks with Gas}
\shortauthors{Roberge \& Weinberger}

\begin{document}

\title{Debris Disks Around Nearby Stars with Circumstellar Gas}

\author{Aki Roberge\altaffilmark{1}}
\affil{Exoplanets and Stellar Astrophysics Laboratory, 
NASA Goddard Space Flight Center, Code 667, Greenbelt, MD, 20771}
\email{Aki.Roberge@nasa.gov}

\and

\author{Alycia J.\ Weinberger}
\affil{Department of Terrestrial Magnetism, 
Carnegie Institution of Washington,  \\ 
5241 Broad Branch Road NW, Washington, DC, 20015}
\email{aweinberger@ciw.edu}

\altaffiltext{1}{NASA Postdoctoral Fellow}

\begin{abstract}

We conducted a survey for infrared excess emission from 16 nearby main sequence shell stars using the Multiband Imaging Photometer for \si\ (MIPS) on the
\emph{Spitzer Space Telescope}.
Shell stars are early-type stars with narrow absorption lines in their spectra that appear to arise from circumstellar (CS) gas.
Four of the 16 stars in our survey showed excess emission at 24~\um\ and 
70~\um\ characteristic of cool CS dust and are likely to be edge-on debris disks. 
Including previously known disks, it appears that the fraction of 
protoplanetary and debris disks among the main sequence shell stars is at 
least $48\% \pm 14\%$.
While dust in debris disks has been extensively studied, relatively little is known about their gas content.
In the case of $\beta$~Pictoris, extensive observations of gaseous species have provided insights into the dynamics of the CS material and surprises about the composition of the CS gas coming from young planetesimals
\citep[e.g.][]{Roberge:2006}. 
To understand the co-evolution of gas and dust through the terrestrial planet formation phase, we need to study the gas in additional debris disks.
The new debris disk candidates from this \si\ survey double the number 
of systems in which the gas can be observed right now with sensitive 
line of sight absorption spectroscopy.  

\end{abstract}




\section{Introduction} \label{sec:intro}

Protoplanetary disks may be classified by their relative abundance of 
primordial gas and dust left over from star formation. 
As the amount of this material decreases, disks evolve from primordial disks 
to transitional disks to debris disks.
Debris disks, which are found around main sequence stars, contain little or no primordial material.
Instead, they are composed of recently produced secondary material generated by collisions between and evaporation of planetesimals, 
analogs of Solar System asteroids and comets. 
Terrestrial planet formation may be occurring in the younger debris disks, while
the older ones seem to correspond to the ``clearing out'' stage early 
in the history of the Solar System, during which most left-over planetesimals are removed from the system. 
Impacts by water-rich planetesimals probably delivered most of the Earth's surface volatiles during the clearing phase \citep[e.g.][]{Morbidelli:2000}.

Gas in debris disks is generally hard to detect.
Only one bona fide debris disk, 49~Ceti, shows any trace of sub-mm CO emission
\citep{Zuckerman:1995, Dent:2005},
indicating that debris disk gas abundances are low relative to those in primordial and transitional disks.
It appears that primordial molecular gas left over from star formation 
has largely dissipated by the debris disk phase.
However, sensitive UV/optical absorption spectroscopy has revealed small 
amounts of secondary gas in a few debris disks: 
$\beta$~Pictoris \citep[e.g.][]{Lagrange:1998}, 
51~Ophiuchi \citep[e.g.][]{Roberge:2002},
$\sigma$~Herculis \citep{Chen:2003}, and
HD~32297 \citep{Redfield:2007a}.
The gas seen is primarily atomic.
The primary production mechanisms for secondary gas in debris disks are currently unknown, but may include photon-stimulated desorption from circumstellar (CS) dust grains \citep{Chen:2007} and/or grain-grain collisions \citep{Czechowski:2007}.

\object[HD39060]{$\beta$~Pic} is the only debris disk whose gas is 
well-characterized.
In this disk, most measured gaseous elements have roughly solar composition relative to each other \citep{Roberge:2006}.
The exception is carbon, which is extremely overabundant relative to every 
other measured element (e.g.\ C$/$O $= 18 \times$ solar).
This discovery was surprising, since the central star has solar 
metallicity \citep{Holweger:1997}.
The carbon overabundance suggests that a previously unsuspected -- and 
currently unknown -- process is operating in the disk, during which the planetesimals preferentially lose volatile carbonaceous material.
However, the lack of similar abundance information for other debris disks renders interpretation of the \bp\ results difficult.  

Detections of gas in debris disks to date have almost exclusively used 
optical or UV absorption spectroscopy of edge-on disks.
Such observations are sensitive to very small amounts of cold gas.  
Unfortunately, the geometric constraint that the line of sight to the 
central star must pass through the disk severely limits the number of disks 
that may be probed for gas in this way.
There is a strong need for additional debris disks in which gas can be observed, in order to study the evolution of gas abundance and composition throughout the entire planet formation phase.

\begin{deluxetable*}{lcccccr}
\centering
\tablecolumns{7}
\tablewidth{0pt}
\tablecaption{Target Stars \label{tab:targets}}
\tablehead{%
\multicolumn{1}{l}{ID} & \colhead{Other Name} & \colhead{Spectral Type} 
& \colhead{Distance} & \colhead{Inside the} 
& \multicolumn{2}{c}{Exp.\ Times (sec)} \\
\colhead{}   & \colhead{}           &  \colhead{}             
& \colhead{(pc)}     & \colhead{LIB?} & \colhead{24~$\mu \mathrm{m}$} 
& \colhead{70~$\mu \mathrm{m}$} }
\startdata
\multicolumn{7}{l}{\textit{Main Sequence Shell Stars from 
\citet{Hauck:2000}}} \vspace*{0.5ex} \\
\object{HD21620} \ \tablenotemark{a} &  	& A0Vn & 143 & y & 37 & 6965 \\
\object{HD24863}		     &  	& A4V  & 107 & y & 37 & 3105 \\
\object{HD39283}       		     & $\xi$~Aur & A2V &  74 & y & 37 & 671 \\
\object{HD77190} \ \tablenotemark{a} & 67~Cnc 	& A8Vn &  59 & y & 37 & 2685 \\
\object{HD98353} \ \tablenotemark{a} & 55~UMa 	& A2V  &  56 & y & 37 & 420 \\
\object{HD118232} 	    	     & 24~Cvn 	& A5V  &  58 & y & 37 & 294 \\
\object{HD142926}	 	     & 4~Her 	& B9pV & 148 & y & 37 & 2853 \\
\object{HD158352}		     &  	& A8V  &  63 & y & 37 & 797 \\
\object{HD196724}		     & 29~Vul  	& A0V  &  65 & y & 37 & 671 \\
\object{HD199603} \ \tablenotemark{a} &  	& A9V  &  85 & y & 37 & 2685 \\
\object{HD223884} 	              &  	& A5V  &  92 & y & 37 & 3483 \\
\object{HD224463} 	   	      &  	& F2V  & 108 & y & 37 & 4280 \\
\vspace*{-1.5ex} \\
\multicolumn{7}{l}{\textit{Shell Stars with Time-Variable Infalling Gas}} 
\vspace*{0.5ex} \\
\object{HD42111} \ \tablenotemark{a}  &  	& A3Vn & 187 & n & 37 & 3609 \\
\object{HD50241}	              & $\alpha$~Pic   & A7IV &  30 & y & 37 
& 168 \\
\object{HD148283} \ \tablenotemark{a} & 25~Her 	& A5V  &  79 & y & 37 & 671 \\
\object{HD217782} \ \tablenotemark{a} & 2~And 	& A3Vn & 107 & y & 37 & 545
\enddata
\tablenotetext{a}{ \ Binary system.} 
\end{deluxetable*}

A class of peculiar stars called shell stars may be the avenue toward 
finding edge-on debris disks containing gas.
These are stars whose spectra show narrow absorption lines arising
from line of sight gas at the velocity of the star. 
Most shell stars are B- or A-type.
In some cases, the narrow lines may arise from interstellar (IS) material, although the fact that the gas is at the velocity of the star makes this interpretation less likely.
In addition, the absorption lines are generally stronger than can be
readily explained by IS gas \citep[e.g.][]{Abt:1973}.
The shell stars have high projected rotational velocities compared 
to non-shell early-type stars, showing that gas is more likely to be 
detected when the line of sight intersects the rotational plane of the system \citep[e.g.][]{Holweger:1999}. 
This indicates that the gas is usually confined to that plane and strengthens the case for a CS rather than IS origin of most shell absorption lines.

Many shell stars appear to be rapidly rotating evolved stars that have 
recently ejected mass from their equatorial planes.
However, \citet{Hauck:2000} evaluated the luminosity classes of 57 shell 
stars and found that 40\% of them are main sequence stars.
The nature of the main sequence shell stars is not fully understood.
Some are rapidly rotating classical Be stars with hot excretion disks. 
We note, however, that 6 out of the 23 main sequence stars in the
\citet{Hauck:2000} sample have protoplanetary or debris disks.
Among the six are the well-known disk systems \bp, HD~163296, and MWC~480.
In fact, \bp\ was classified as a shell star a decade before discovery of its dust disk \citep{Slettebak:1975}.

In this paper, we present a study aimed at finding additional debris disks hiding among the shell stars.
We used the Multiband Imaging Photometer for \si\ (MIPS) on the \emph{Spitzer Space Telescope} to measure the fluxes from 16 nearby main sequence shell stars at 24 and 70 \um.
The observations were designed to probe sensitively for excess flux arising 
from CS dust.
The goal is to assemble a unique and valuable set of debris disk systems 
where the gas may be studied in as much detail as the dust. 

\section{Target Selection and Observations} \label{sec:obs}


We began our target selection by considering which subset of the 57 shell stars 
in \citet{Hauck:2000} is the most likely to contain debris disks.
To eliminate evolved stars, we selected only the stars with luminosity class~V
(23 stars).
We then eliminated six stars already known to have protoplanetary or debris disks. 
We also wanted to eliminate ordinary stars that might have been misclassified as shell stars simply because they have anomalously large amounts of IS material along their lines of sight.
Therefore, we discarded stars that are outside of the Local Interstellar Bubble (LIB), using the LIB size and shape from \citet{Lallement:2003}.
This left us with 12 stars (listed in Table~\ref{tab:targets}).

To this set, we added four shell stars whose spectra show narrow, time-variable absorption features.
%
%
These are HD~42111 \citep{LecavelierdesEtangs:1997a}, 
HD~50241 \citep{Hempel:2003}, HD~148283 \citep{Grady:1996},
and HD~217782 \citep{Cheng:2003}.
The time-variability confirms that the gas along the lines of sight to these stars is CS, despite the fact that one of them is outside the LIB.
Such behavior is seen in spectra of \bp, and is caused by evaporation 
of star-grazing planetesimals passing through the line of sight 
\citep[e.g.][]{Beust:1990}.
One of the four stars (HD~50241) has luminosity class~IV.
However, since it has time-variable \textit{infalling} CS gas, it is 
unlikely to be an evolved mass-losing star.

We obtained MIPS 24~\um\ images and 70~\um\ default-scale images for 15 of 
our 16 target stars.
For one of our stars (HD~77190), we obtained only a 70~\um\ observation.
This star was observed at 24~\um\ in GTO program 40; we used that image in our study.
All the data were calibrated with version S13.2.0 of the \si\ Science Center
(SSC) calibration pipeline software. 
The total exposure times for each star are listed in the last two columns of 
Table~\ref{tab:targets}.

\section{Analysis} \label{sec:anal}

\begin{deluxetable*}{lrrrcl}
\centering
\tablecolumns{6}
\tablewidth{0pt}
\tablecaption{Photometry Apertures and Aperture Corrections 
\label{tab:aper_corr}}
\tablehead{%
\colhead{Wave} & \colhead{$r_\mathrm{object}$} 
& \colhead{$r_\mathrm{sky \ inner}$} 
& \colhead{$r_\mathrm{sky \ outer}$} 
& \colhead{Aperture} & \colhead{Comment} \\ 
\colhead{(\um)} & \colhead{} & \colhead{} & \colhead{} 
& \colhead{Correction} & \colhead{} }
\startdata
24	& 14\farcs7 & 29\farcs4 & 41\farcs7 & 1.143 & default 
\tablenotemark{a} \\
24 & 14\farcs7 & 39\farcs2 & 49\farcs0 & 1.135 & used for HD~42111 
\tablenotemark{b} \\
24 & 14\farcs7 & 19\farcs6 & 29\farcs4 & 1.154 & used for HD~158352 
\tablenotemark{b} \\
24 &  9\farcs8 & 19\farcs6 & 31\farcs9 & 1.322 & used for HD~199603 
\tablenotemark{b} \\
70 & 28\farcs0 & 40\farcs0 & 68\farcs0 & 1.298 & default 
\tablenotemark{a} \\
70 & 16\farcs0 & 40\farcs0 & 64\farcs0 & 1.884 & used for HD~39283, HD~77190, HD~42111, \& HD~158352 \tablenotemark{c} \\
70 &  8\farcs0 & 40\farcs0 & 64\farcs0 & 4.186 & used for HD~199603 
\tablenotemark{c} 
\enddata
\tablenotetext{a}{ \ From \citet{Su:2006}.}
\tablenotetext{b}{ \ Calculated from our data. See \S\ref{sub:24}.}
\tablenotetext{c}{ \ From the SSC website (http://ssc.spitzer.caltech.edu/mips/apercorr/).}
\end{deluxetable*}

\subsection{24~\um\ Photometry} \label{sub:24}

To determine the 24~\um\ fluxes from the target stars, we used the 
pipeline-processed post-Basic Calibrated Data (BCD) mosaic images, which 
are suitable for detailed analysis.
The pixels in these images are $2\farcs45 \times 2\farcs45$.
Every target star was detected at $S/N \geq 50$, except for the faintest (HD~224463), which was detected at $S/N = 7$.
The locations of the star centers were determined by fitting a 2-D Gaussian
to each star. 

For most of the target stars, we performed aperture photometry using an object aperture and sky annulus combination close to the 24~\um\ large aperture combination used in \citet{Su:2006} ($r_\mathrm{obj} = 14\farcs7, \ r_\mathrm{sky \ inner} = 29\farcs4, \ r_\mathrm{sky \ outer} = 41\farcs7$).
The aperture correction for this combination is 1.143.
For three of the targets, there are nearby faint stars or background objects that fall within the default object aperture or sky annulus.
In these cases, we used aperture and annulus combinations that avoid the nearby sources.
Aperture corrections for these combinations were calculated using the 24~\um\ image of our brightest target star at this wavelength (HD~50241).
To verify the accuracy of our calculated corrections, we used that image to calculate a correction for the \citet{Su:2006} aperture combination given above.
Our calculated correction differed from the published \citet{Su:2006} correction by only 0.5\%.
All the apertures, annuli, and corrections used appear in 
Table~\ref{tab:aper_corr}.

\begin{deluxetable*}{lrr@{\extracolsep{7pt}}c@{\extracolsep{-2pt}}r%
@{\extracolsep{-3pt}}crrr@{\extracolsep{-2pt}}r@{\extracolsep{-3pt}}c}
%
%
\centering
\tablecolumns{11}
\tablewidth{0pt}
\tablecaption{Photometry Results \label{tab:results}}
\tablehead{%
\multicolumn{1}{l}{ID} & \multicolumn{5}{c}{24~\um} 
& \multicolumn{5}{c}{70~\um} \\
\cmidrule(lr){2-6}\cmidrule(lr){7-11}
\colhead{} & \colhead{$F_\star$} & \colhead{$F_\mathrm{MIPS}$} 
& \colhead{$\sigma_\mathrm{MIPS}$} 
& \colhead{$\left( (F_\mathrm{MIPS} - F_\star) / F_\star \right) * 100$} 
& \colhead{Excess?} 
& \colhead{$F_\star$} & \colhead{$F_\mathrm{MIPS}$} 
& \colhead{$\sigma_\mathrm{MIPS}$} 
& \colhead{$\left( (F_\mathrm{MIPS} - F_\star) / F_\star \right) * 100$} & \colhead{Excess?} \\
\colhead{} & \colhead{(mJy)} & \colhead{(mJy)} & \colhead{(mJy)} 
& \colhead{(\%)} & \colhead{} & \colhead{(mJy)} & \colhead{(mJy)} 
& \colhead{(mJy)} & \colhead{(\%)} & \colhead{} } 
\startdata
HD21620 &  27.31 & $ 34.11$ & $ 0.64$ & $  24.88 \pm  2.34$ \hspace*{6ex} & yes &   3.3 & $ 24.2$ & $  6.5$ \hspace*{1ex} & $ 640.1 \pm 199.1$ \hspace*{5ex} & yes \\
HD24863 &  25.06 & $ 28.01$ & $ 0.55$ & $  11.77 \pm  2.20$ \hspace*{6ex} & no &   3.0 & $<  19.4$ & $  5.3$ \hspace*{1ex} & $<  545.4$ \hspace*{8ex} & no \\
HD39283 &  84.22 & $ 84.88$ & $ 0.70$ & $   0.79 \pm  0.83$ \hspace*{6ex} & no &  10.1 & $<  23.5$ & $  5.7$ \hspace*{1ex} & $<  133.9$ \hspace*{8ex} & no \\
HD77190 &  45.08 & $ 48.81$ & $ 0.50$ & $   8.29 \pm  1.12$ \hspace*{6ex} & no &   5.4 & $<  16.2$ & $  3.5$ \hspace*{1ex} & $<  200.1$ \hspace*{8ex} & no \\
HD98353 & 108.95 & $114.52$ & $ 0.53$ & $   5.11 \pm  0.49$ \hspace*{6ex} & no &  13.0 & $<  54.1$ & $ 12.3$ \hspace*{1ex} & $<  315.3$ \hspace*{8ex} & no \\
HD118232 & 111.10 & $155.67$ & $ 0.49$ & $  40.12 \pm  0.44$ \hspace*{6ex} & yes &  13.3 & $ 81.4$ & $  7.6$ \hspace*{1ex} & $ 513.7 \pm  57.2$ \hspace*{5ex} & yes \\
HD142926 &  28.55 & $107.52$ & $ 0.51$ & $ 276.66 \pm  1.79$ \hspace*{6ex} & yes &   3.4 & $ 25.4$ & $  6.3$ \hspace*{1ex} & $ 651.6 \pm 186.8$ \hspace*{5ex} & yes \\
HD158352 &  74.02 & $103.44$ & $ 0.35$ & $  39.75 \pm  0.47$ \hspace*{6ex} & yes &   8.9 & $235.1$ & $  7.3$ \hspace*{1ex} & $2551.2 \pm  82.8$ \hspace*{5ex} & yes \\
HD196724 &  73.44 & $ 75.21$ & $ 0.46$ & $   2.41 \pm  0.63$ \hspace*{6ex} & no &   8.7 & $ 17.2$ & $  4.9$ \hspace*{1ex} & $  97.0 \pm  56.2$ \hspace*{5ex} & no \\
HD199603 &  44.71 & $ 50.54$ & $ 0.38$ & $  13.04 \pm  0.85$ \hspace*{6ex} & no &   5.4 & $ 18.2$ & $  3.8$ \hspace*{1ex} & $ 239.7 \pm  71.0$ \hspace*{5ex} & yes $^{a}$ \\
HD223884 &  34.45 & $ 37.23$ & $ 0.39$ & $   8.06 \pm  1.13$ \hspace*{6ex} & no &   4.1 & $<  21.1$ & $  4.5$ \hspace*{1ex} & $<  412.2$ \hspace*{8ex} & no \\
HD224463 &   3.79 & $  4.07$ & $ 0.59$ & $   7.50 \pm 15.51$ \hspace*{6ex} & no &   0.5 & $<  11.7$ & $  3.6$ \hspace*{1ex} & $< 2469.1$ \hspace*{8ex} & no \\
HD42111 &  47.02 & $ 51.13$ & $ 0.74$ & $   8.73 \pm  1.56$ \hspace*{6ex} & no &   5.6 & $<   9.0$ & $  3.0$ \hspace*{1ex} & $<   59.7$ \hspace*{8ex} & no \\
HD50241 & 608.29 & $647.56$ & $ 0.42$ & $   6.46 \pm  0.07$ \hspace*{6ex} & no &  73.0 & $ 75.9$ & $  9.7$ \hspace*{1ex} & $   4.0 \pm  13.3$ \hspace*{5ex} & no \\
HD148283 &  66.95 & $ 71.03$ & $ 0.43$ & $   6.10 \pm  0.64$ \hspace*{6ex} & no &   8.0 & $<  21.2$ & $  5.2$ \hspace*{1ex} & $<  164.4$ \hspace*{8ex} & no \\
HD217782 &  88.28 & $ 92.10$ & $ 0.56$ & $   4.33 \pm  0.63$ \hspace*{6ex} & no &  10.6 & $<  41.9$ & $ 10.4$ \hspace*{1ex} & $<  296.8$ \hspace*{8ex} & no \\

\enddata
\tablenotetext{a}{ \ This ``excess'' is probably caused by a nearby source.}
\end{deluxetable*}

The results of our aperture photometry, which have not been color-corrected, appear in Table~\ref{tab:results}. 
The uncertainties in the flux measurements were determined in the following
way. 
The aperture photometry equation giving the fluxes in mJy is 
\begin{equation}
F = c \: A \: p \left( \sum_{i, j} f_{i, j} - 
N * \overline{s}  \right) \: ,
\end{equation}
where $c$ is a units conversion factor (0.023504 mJy sr arcsec$^{-2}$ 
MJy$^{-1}$), $A$ is the aperture correction, $p$ is the angular size of the pixels (2.45$^2$ arcsec$^2$), $f_{i, \, j}$ is the surface brightness value of pixel $i, j$ in units of MJy sr$^{-1}$, $N$ is the number of pixels in the object aperture, and $\overline{s}$ is the mean sky brightness calculated from the pixels in the sky annulus. 
The sum is over all the pixels in the object aperture.
This equation may be reduced to 
\begin{equation}
F = C \left( \sum_{i, j} f_{i, j} - B \right) \: ,
\end{equation}
where $C$ is a constant and $B$ is the estimate of the background flux in the object aperture. 
Propagating uncertainties gives the following equation for the uncertainty in the total flux.
\begin{equation}
\sigma_F = C * \sqrt{ \sum_{i, \, j} \sigma_{s_{i, \, j}}^2 + \sigma_{B}^2} \: ,
\end{equation}
where the $\sigma_{s_{i, \, j}}$ are the uncertainties in the surface brightness values and $\sigma_{B}$ is the uncertainty in the estimate of the background flux in the object aperture.

The $\sigma_{s_{i, \, j}}$ values were taken from the uncertainty array provided by the SSC calibration pipeline. 
To estimate $\sigma_{B}$, we placed eight apertures and sky annuli in empty regions around each star.
For each empty aperture, the difference between the actual value of $B$
and the value estimated from the pixels in the sky annulus was calculated.
The standard deviation of the differences was taken as the estimate of 
$\sigma_{B}$.
This method of calculating the flux uncertainties includes object noise, 
noise due to variations of the background within the sky annulus, and
noise due to spatial variations of the background between the object aperture and the sky annulus. 

We calculated the median background surface brightness in each image using pixels between 61\arcsec\ and 135\arcsec\ from the target star. 
The background values range from 17.3~MJy/sr (low) to 49.7~MJy/sr (medium-high).
The definitions of low, medium, and high background levels may be found 
on the SSC website\footnote{http://ssc.spitzer.caltech.edu/obs/bg.html}.
For our target stars, variation of the background between the object aperture and the sky annulus appears to be responsible for 18\% to 67\% of the total flux uncertainty.

\subsection{70~\um\ Photometry} \label{sub:70}

\begin{figure}[b!]
\centering
\plotone{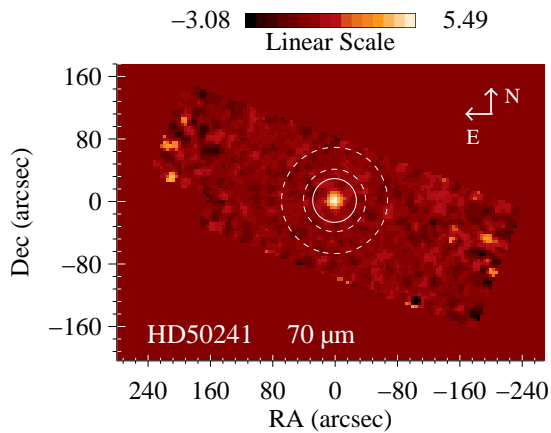}
\caption{MIPS 70~\um\ image of HD~50241.  The mosaic was created using MOPEX.  The pixels are $4\farcs0 \times 4\farcs0$. The default object
aperture used for the 70~\um\ photometry is overlaid with a solid white line.
The default sky annulus is shown with dashed white lines. \label{fig:HD50241} }
\end{figure}

Since the 70~\um\ post-BCD mosaic images are not suitable for detailed 
analysis, we used the SSC software MOPEX to combine the filtered BCD images. 
MOPEX was set to propagate the uncertainty images from the SSC pipeline, 
and to use multiframe temporal outlier rejection, since all the mosaics have good coverage values ($>10$ exposures in the regions around the target stars).
The pixels in the final mosaics are $4\farcs0 \times 4\farcs0$. 
One of the final 70~\um\ mosaics is shown in Figure~\ref{fig:HD50241}.
Seven of the 16 target stars were detected at 70~\um\ with $S/N \geq 3$.
Since many of the stars were not detected, we used the telescope pointing information in the image headers to determine the locations of the star centers. 

For most of the stars, we used an object aperture and sky annulus combination close to the 70~\um\ large aperture combination used in \citet{Su:2006} 
($r_\mathrm{obj} = 28\farcs0, \ r_\mathrm{sky \ inner} = 40\farcs0, \ r_\mathrm{sky \ outer} = 68\farcs0$).
The default object aperture and sky annulus are overlaid on the 70~\um\ mosaic shown in Figure~\ref{fig:HD50241}.
For stars with nearby faint sources in the 24~\um\ images, we used aperture and annulus combinations that avoided the sources, even if they were not visible in the 70~\um\ images.
The aperture corrections for these combinations were taken from tables available on the SSC website\footnote{http://ssc.spitzer.caltech.edu/mips/apercorr/}.
All the apertures, annuli, and corrections used appear in 
Table~\ref{tab:aper_corr}. 
The 70~\um\ photometry results, which have not been color-corrected, appear in Table~\ref{tab:results}.

We calculated the uncertainties in the 70~\um\ fluxes following the same procedure used for the 24~\um\ photometry (described in \S\ref{sub:24}).
For each star not detected at the $3 \sigma$ level, a conservative upper limit 
on the 70~\um\ flux was set by adding 3 times the total statistical uncertainty to any positive flux measured in the object aperture. 
The median background surface brightnesses near the target stars were estimated from the unfiltered post-BCD mosaic images, using pixels between 48$\arcsec$ 
and 80$\arcsec$ from the stars. 
The background values range from 5.8~MJy/sr (low) to 16.3~MJy/sr (medium).
At 70~\um, background variation between the object aperture and the sky annulus appears to be responsible for 43\% to 84\% of the total flux uncertainty 
for our target stars. 

\section{Spectral Energy Distributions} \label{sec:seds}

\subsection{Fitting Photospheric Models} \label{sub:stellar}

To see if any of the target stars show excess IR emission, we first needed to determine their expected photospheric fluxes at 24~\um\ and 70~\um\ for comparison to the observed fluxes. 
This was done by fitting stellar models to the optical and near-IR fluxes 
from the stars.
$B$ and $V$ magnitudes from the Tycho-2 Catalogue (H{\o}g et al.\ 2000) 
\nocite{Hog:2000} were converted to Johnson magnitudes using formulae given
in Volume~1 of the Hipparcos and Tycho Catalogues\footnote{http://www.rssd.esa.int/Hipparcos/CATALOGUE\_VOL1/sect1\_03.pdf}, 
then to flux densities.
The near-IR magnitudes in the $J$, $H$, and $K_\mathrm{s}$ bands from the 
2MASS Catalog were converted to flux densities using the zero-points given in the 2MASS Explanatory Supplement\footnote{http://www.ipac.caltech.edu/2mass/releases/allsky/doc/sec6\_4a.html}. 

Since all but one of the target stars are within the LIB, we did not 
apply an IS extinction correction to the optical and near-IR photometry.
Kurucz model photospheric spectra, with $\log g = 4.0$ for the stars with spectral types earlier than A5 and $\log g = 4.5$ for the cooler stars,
were fit to the optical and near-IR fluxes using $\chi^2$ minimization. 
The best-fitting model spectrum for each star was used to calculate the 
predicted photospheric fluxes at 24~\um\ and 70~\um, which appear in 
Table~\ref{tab:results}.

We then considered whether any of the stars showed significant excess flux 
at either wavelength. 
The excesses were characterized by calculating the percent deviation of the measured flux from the predicted photospheric flux, while propagating the statistical uncertainties in the measured fluxes (see Table~\ref{tab:results}).
First, we compared the excesses to the uncertainties in the absolute 
photometric calibration of MIPS data processed with version S13 of the SSC calibration pipeline.
The uncertainties are 4\% for 24~\um\ data and 7\% for 70~\um\ data 
(MIPS Data Handbook, version~3.2).
We conservatively designated excesses greater than 5 times these uncertainties as significant, which corresponds to $>20\%$ at 24~\um\ and $>35\%$ at 70~\um.
The second criterion for a significant excess is that it must be at least 
3 times greater than its propagated statistical uncertainty.
This criterion eliminated apparent excesses only at 70~\um. 

Four of the 16 stars showed significant 24~\um\ excesses 
(HD~21620, HD~118232, HD~142926, and HD~158352).
These stars also showed significant 70~\um\ excesses.
The spectral energy distributions (SEDs) of all the target stars are shown 
in Figure~\ref{fig:seds}.
One additional star, HD~199603, showed a significant 70~\um\ excess.
However, the source visible in the 70~\um\ image of HD~199603 is slightly 
offset from the expected position of the A-star;
it is located at the position of a nearby faint source visible in the 24~\um\ image of HD~199603.
Therefore, we believe that the apparent 70~\um\ excess from HD~199603 is due to contamination from this nearby red source, despite our care in aperture selection.
The nature of this nearby red source is not yet known.

\begin{figure*}
\centering
\plotone{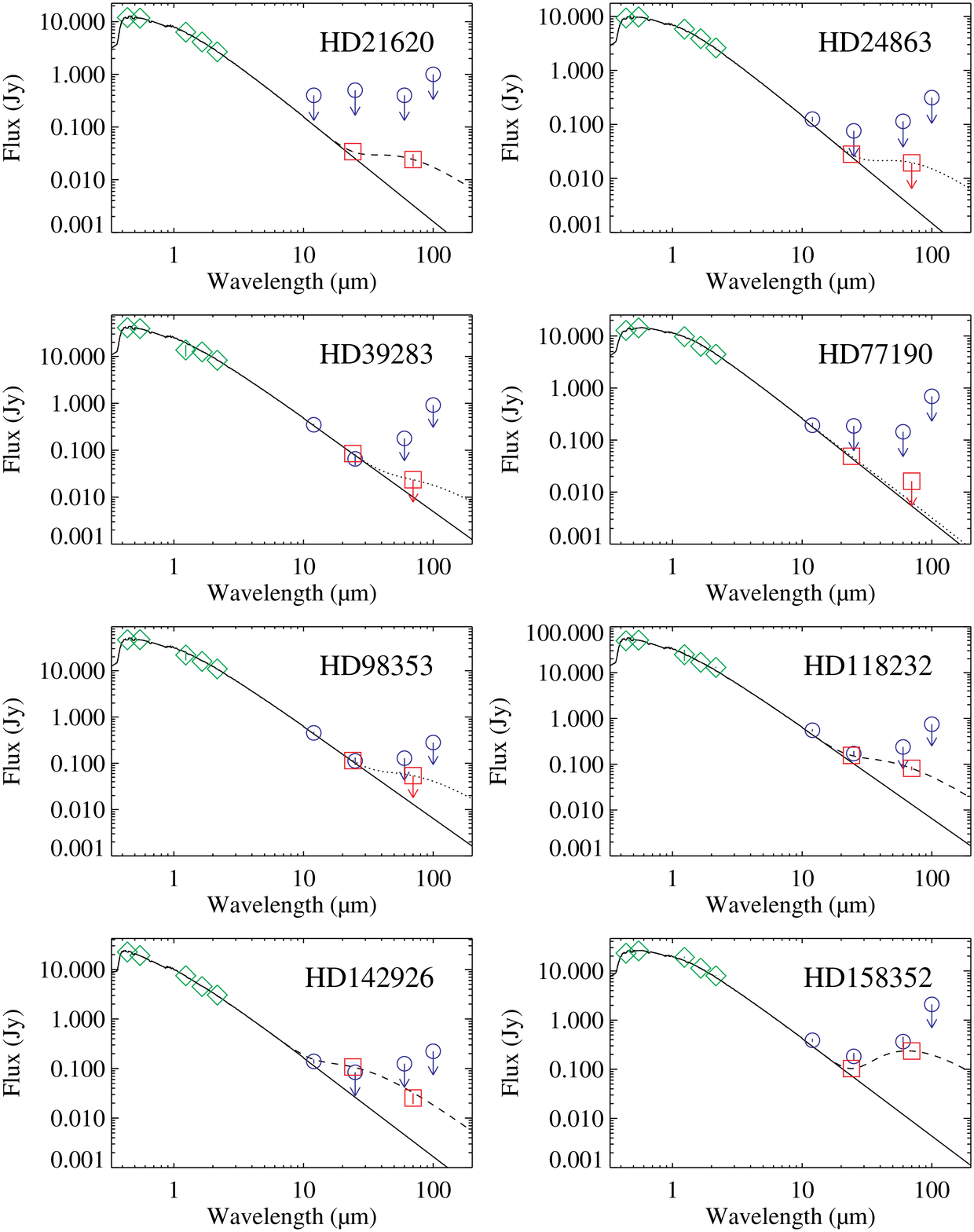}
\caption{SEDs of the target stars.
The fluxes in the $B$ and $V$ bands from the Tycho-2 Catalogue and the $J$, $H$, and $K_\mathrm{s}$ bands from the 2MASS Catalog are plotted with green diamonds.
Color-corrected fluxes from \iras\ are plotted with blue circles. 
The new 24~\um\ and 70~\um\ MIPS fluxes, which have not been color corrected, are plotted with red squares.
Upper limits are indicated with arrows.
Uncertainties in the flux measurements are overplotted with straight vertical lines.
The best-fitting Kurucz model photospheric spectra are overplotted with 
solid black lines.
In each panel, the total SED model (the sum of the stellar model and a 
single-temperature blackbody model) is shown with dashed or dotted black lines.
The blackbody temperatures and \lir\ values for these models are given in 
Table~\ref{tab:seds}.
For the four stars with significant IR excess emission at the MIPS wavelengths
(HD~21620, HD~118232, HD~142926, \& HD~158352),
the best-fitting total SED model is shown with a dashed line. 
For the stars without significant IR excesses, the total SED model showing the upper limit on the IR excess emission is overplotted with a dotted line.
\label{fig:seds} }
\end{figure*}

\begin{figure*}
\centering
\figurenum{2}
\plotone{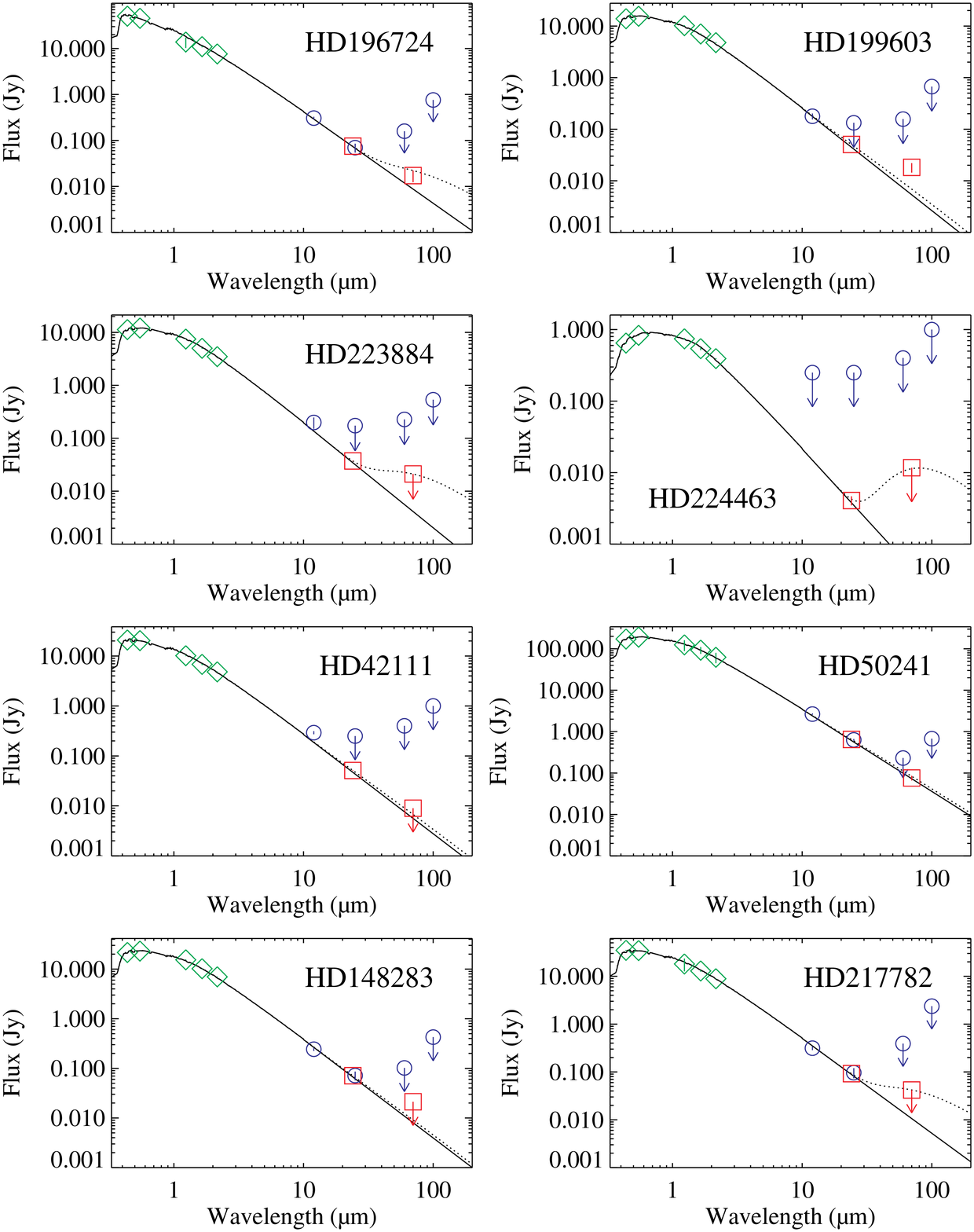}
\caption{Continued.}
\end{figure*}

For the rest of the target stars, the mean 24~\um\ excess is $6.3\% \pm 3.1\%$.  
There is a small positive offset, although it is not significant at the 
$3\sigma$ level.  
Ordinarily, this would indicate a problem with the photospheric modeling or 
with the aperture photometry itself.  
However, we do not believe this to be case. 
The photometry should be accurate, since the MIPS color corrections for 
early-type stars are extremely small.
The photospheric modeling appears accurate.
The average deviation between the fitted effective temperatures and the 
expected effective temperatures (based on the spectral types) is only 378~K.
The accuracy of the fitted effective temperatures also indicates that IS reddening does not significantly affect the photospheric model fitting.

The small positive offset measured at 24~\um\ may be real.
None of the target stars are unambiguously non-excess stars, since they are 
all main sequence shell stars with CS gas.
If they are not debris disks or protoplanetary disks, then they may be 
classical Be stars, which show small, hot mid-IR excesses from free-free emission \citep[see Figure~6 in][]{Su:2006}.
We plan additional studies to clarify the nature of the target shell stars without significant MIPS excesses.

\subsection{Fitting Models to Excess Emission} \label{sub:dustmod}

The next steps were to begin characterization of the CS material responsible 
for the detected excesses and to set upper limits on the amount of material around the stars without significant excesses.
We modeled the total SED of each star as the sum of the best-fitting photospheric model spectrum and a single-temperature blackbody dust emission model.
The free parameters of the total SED model were the blackbody dust temperature
($T_\mathrm{BB}$) and the fractional IR luminosity (\lir). 
The model was fit to the MIPS 24 and 70~\um\ fluxes using $\chi^2$ minimization.
If the star's 12~\um\ flux from the \textit{Infrared Astronomical Satellite} (\iras) was a detection, it was included in the fitting. 
\iras\ upper limits and detections at other wavelengths were not included; 
the \iras\ photometry has large upper limits and measurement uncertainties compared to the MIPS photometry and did not usefully constrain the fit. 

\begin{deluxetable*}{lcrcc}
\centering
\tablecolumns{5}
\tablewidth{0pt}
\tablecaption{Results from SED Fitting \label{tab:seds}}
\tablehead{%
\multicolumn{1}{l}{ID} & \colhead{$T_\mathrm{BB}$} 
& \multicolumn{1}{c}{$L_\mathrm{IR}/L_\star$} 
& \multicolumn{2}{c}{Dust Distance \ (AU)} \\
\colhead{} & \colhead{(K)} & \colhead{} 
& \colhead{BB \tablenotemark{a}} & \colhead{Silicates \tablenotemark{b} } }
\startdata
HD21620 	& $93.4 \pm 8.2$	& $(2.51 \pm 0.49) \times 10^{-5}$ & 63 & 174 \\
HD24863 	& \nodata 		& $< 1.9 \times 10^{-5}$ & &  \\
HD39283 	& \nodata 		& $< 3.1 \times 10^{-6}$ & &  \\
HD77190 	& \nodata 		& $< 1.3 \times 10^{-5}$ & &  \\
HD98353 	& \nodata   		& $< 9.5 \times 10^{-6}$ & &  \\
HD118232 	& $114.0 \pm 4.4$ 	& $(2.56 \pm 0.12) \times 10^{-5}$ & 36 & 89 \\
HD142926 	& $221.6 \pm 19.6$ 	& $(8.67 \pm 0.84) \times 10^{-5}$ & 18 & 44 \\
HD158352 	& $76.1 \pm 0.79$ 	& $(9.29 \pm 0.26) \times 10^{-5}$ & 62 & 191 \\
HD196724 	& \nodata 		& $< 2.4 \times 10^{-6}$ & &  \\
HD199603 	& \nodata 		& $< 1.8 \times 10^{-5}$ & &  \\
HD223884 	& \nodata 		& $< 1.6 \times 10^{-5}$ & &  \\
HD224463 	& \nodata 		& $< 1.1 \times 10^{-4}$ & &  \\
HD42111 	& \nodata 		& $< 1.0 \times 01^{-5}$ & &  \\
HD50241 	& \nodata 		& $< 9.8 \times 10^{-6}$ & &  \\
HD148283 	& \nodata 		& $< 8.4 \times 10^{-6}$ & &  \\
HD217782 	& \nodata 		& $< 9.6 \times 10^{-6}$ & &  
\enddata
\tablecomments{The upper limits on \lir\ are valid for values of
$T_\mathrm{BB}$ between 50~K and 300~K.} 
\tablenotetext{a}{ \ Calculated assuming blackbody grains.}
\tablenotetext{b}{ \ Calculated assuming 1~\um-size astronomical silicate grains.}
\end{deluxetable*}

For the four stars with significant IR excess, the dust parameters of the 
best-fitting SED models appear in Table~\ref{tab:seds}.
We estimated the uncertainties in these parameters in the following way. 
We compared total SED models, each of which was the sum of a Kurucz photospheric spectrum model and a single-temperature blackbody emission model, to the optical, near-IR, and mid-IR photometry simultaneously.
All the stellar and blackbody model parameters were varied and $\chi^2$ 
values calculated over large regular grids of parameter space 
(which could be described as the ``brute force'' method). 
This is in contrast to the fitting procedure described in the previous 
paragraph, which used the downhill simplex method to rapidly find the 
$\chi^2$ minimum. 

The contours of $\chi^2$ from the brute force minimization were used to determine the $\pm 1 \sigma$ uncertainties on the model parameters
shown in Table~\ref{tab:seds}.
This method should give accurate estimates of the uncertainties even if any model parameter is correlated with any other. 
The contours showed that the stellar parameters ($T_\mathrm{eff}$ and a normalization factor) are correlated with each other, as are the blackbody
parameters ($T_\mathrm{BB}$ and \lir).
As expected, the stellar parameters are nearly independent of the blackbody parameters.
This is why the parameters of the best-fitting model from each brute force 
$\chi^2$ minimization agree with those found using the method described in the first paragraph of this subsection, in which minimization was done first for the stellar model and afterwards for the blackbody model. 

We next determined upper limits on \lir\ for the 12 stars without significant 
24 or 70~\um\ excesses.
The procedure was similar to that described in the first paragraph of this subsection.
For each star, we fit the sum of the best-fitting Kurucz model spectrum and blackbody emission models with fixed values of $T_\mathrm{BB}$ to the 
\iras\ 12~\um\ flux if it was not an upper limit and the MIPS 24 and 70~\um\ fluxes even if they were upper limits. 
The $\chi^2$ minimization code used forced the models to 
1) stay below upper limits, 
2) stay below $F_\mathrm{MIPS} + 1 \sigma$ for detections, and 
3) stay above the photospheric flux.
The best-fitting models with fixed values of $T_\mathrm{BB}$ between
20~K and 300~K were determined.
For each of these models, the \lir\ value is the upper limit assuming
that particular blackbody temperature. 

\begin{figure*}
\centering
\epsfig{file=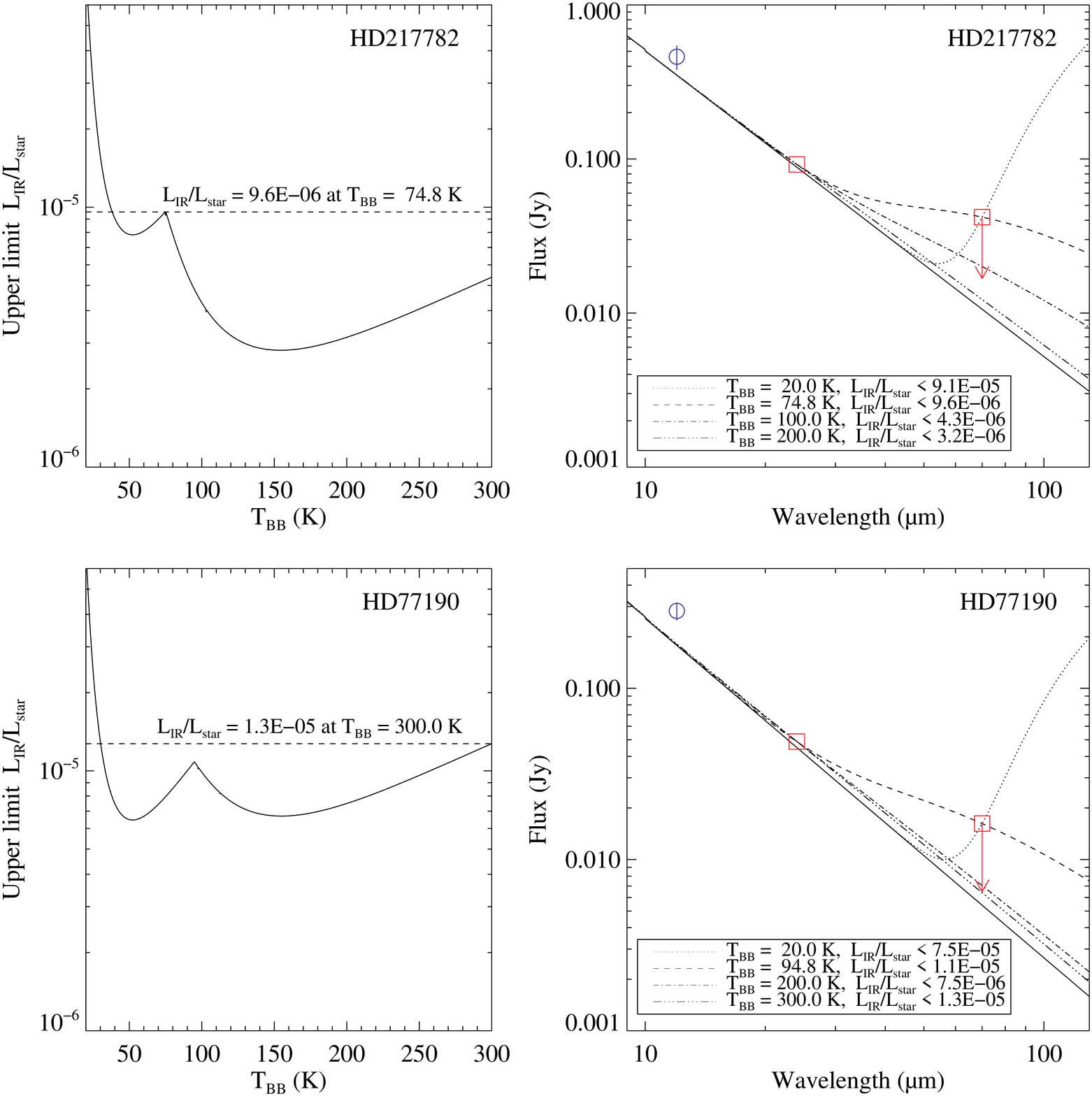,width=6.0in}
\caption{\lir\ upper limits as a function of blackbody temperature.
In the left-hand panels, \lir\ upper limits for HD~217782 and HD~77190 are plotted vs.\ $T_\mathrm{BB}$.
The largest \lir\ upper limits for 50~K $< T_\mathrm{BB} <$ 300~K are
indicated with horizontal dashed lines. 
In the right-hand panels, portions of the SEDs of these two stars are shown. 
The best-fitting upper limit SED models for four values of $T_\mathrm{BB}$ are overplotted with dotted, dashed, and dot-dashed lines.
The color-corrected \iras\ 12~\um\ flux is plotted with a blue circle.
The MIPS 24~\um\ and 70~\um\ fluxes are plotted with red squares; 
70~\um\ upper limits are indicated by arrows. 
See \S\ref{sub:dustmod} for a discussion of this figure. \label{fig:upper} }
\end{figure*}

Figure~\ref{fig:upper} illustrates the general behavior of the \lir\ upper limits as a function of $T_\mathrm{BB}$.
Plots of the \lir\ upper limits vs.\ $T_\mathrm{BB}$ for HD~217782 and
HD~77190 appear in the panels on the left.
The panels on the right show portions of the two stars' SEDs, with the 
best-fitting upper limit SED models for four values of $T_\mathrm{BB}$ overplotted.
The \lir\ upper limits decrease sharply with increasing temperature between 
20~K and about 50~K.
This is because we do not have photometry at longer wavelengths, which would constrain the fitting of SEDs with cold blackbody temperatures. 
As $T_\mathrm{BB}$ increases, the \lir\ upper limits increase then decrease, 
forming an absolute maximum (e.g.\ HD~217782) or local maximum (e.g.\ HD~77190) somewhere between $\sim 50$~K and $\sim 150$~K.
The \lir\ upper limits then gradually increase again.
The maximum between $\sim 50$~K and $\sim 150$~K occurs at the temperature 
of the best-fitting SED model found if one treats a 70~\um\ upper limit as a detection (see the right-hand panels of Figure~\ref{fig:upper}).

However, for five of the stars without significant excesses (including HD~77190), the largest \lir\ upper limit occurs at 300~K (see the bottom panels of Figure~\ref{fig:upper}).
In these cases, SED models with fairly high blackbody temperatures are consistent with the limits imposed by the 12~\um\ and 24~\um\ fluxes, 
indicating that the SED fitting is also poorly constrained at shorter mid-IR wavelengths. 
The results of this analysis are shown in Table~\ref{tab:seds}.
For the stars without significant IR excess, the \lir\ upper limit given is valid for all blackbody temperatures between 50~K and 300~K.

\section{Discussion} \label{sec:disc}

\subsection{Stars With Excesses} \label{sub:excess}

This first survey for debris disks around main sequence shell stars was successful;
4 out of the 16 stars we observed with MIPS show IR excess emission at 
24~\um\ and 70~\um.
The dust temperatures from simple modeling of their SEDs are cool, 
similar to those of other debris disks. 
These candidate debris disks have best-fitting \lir\ values that are about 
30 to 100 times lower than that of \bp.
The mid-IR excesses of these four stars do not resemble those of classical Be stars, whose mid-IR fluxes observed with \si\ show a steep power-law decline with increasing wavelength \citep[see Figure~6 in][]{Su:2006}.

For the stars with excesses, we calculated the distances from the central stars of grains with these temperatures, assuming blackbody grains and 1~\um-size 
astronomical silicate grains. 
The distance calculated assuming blackbody grains represents a minimum 
distance, while the one calculated assuming silicate grains is probably more accurate. 
The values appear in Table~\ref{tab:seds}.
For HD~21620, HD~118232, and HD~158352, the calculated distances suggest that their CS dust lies in the Kuiper Belt region for an A-star \citep{Su:2006}.
The dust around HD~142926 is somewhat warmer, and may lie closer to the 
central star. 

We have searched the literature for information on the ages of our debris disk
candidates. 
HD~21620 is a member of the $\alpha$~Per open cluster, which indicates that it is about 80~Myr old.
This age is fairly young compared to most debris disks, whose ages 
range from $\approx 10$~Myr (e.g.\ \bp) to more than 1~Gyr.
We were not able to find any published age estimate for HD~118232.
In a future paper, we will estimate the age of this star using multiple 
age-dating techniques.

An estimate of the age of HD~142926 appears in \citet{Zorec:2005}.
In this paper, the ages of Be stars were determined by comparing their positions on the H-R diagram to theoretical tracks of the evolution of rapidly rotating stars from the zero-age main sequence (ZAMS) to the terminal-age main sequence (TAMS) and beyond into the post-main sequence phases. 
\citet{Zorec:2005} estimate the age of HD~142926 to be about 78~Myr old, 
which is $\approx 37\%$ of the star's total main sequence lifetime. 

Based on the \iras\ fluxes from HD~158352, \citet{Oudmaijer:1992} 
tentatively suggested that it has an IR excess.
A new analysis of the \iras\ data by M{\'o}or et al.\ (2006) 
\nocite{Moor:2006} reaffirmed this suggestion, which is now confirmed by 
our MIPS photometry.
HD~158352 is not known to be a member of any young moving group.
Its age is estimated to be $750^{+150}_{-150}$~Myr, based on the star's 
location on the H-R diagram relative to theoretical isochrones 
(M{\'o}or et al.\ 2006). \nocite{Moor:2006}
The IR excess from this star is anomalously large given its presumed advanced age. In a recent \si\ MIPS survey for excesses from nearby main sequence early-type stars, there are no stars older than about 300~Myr that show as large a 
24~\um\ excess as HD~158352 \citep{Su:2006}.

\subsection{Stars Without Excesses} \label{sub:noexcess}

We here discuss two particular stars from our survey that do not have significant IR excesses.
HD~50241 ($\alpha$~Pic) does not show significant excess emission at either MIPS wavelength, although a claim for IR excess from this star has been made based 
on its fluxes from \iras\ \citep{Song:2001}.
This star is often used as a point-spread function reference for coronagraphic imaging of the \bp\ debris disk \citep[e.g.][]{Golimowski:2006}, so it is 
fortunate that it does not have a large amount of CS dust. 

There has been a claim for star-grazing planetesimals around HD~217782
\citep{Cheng:2003}, which would make it a true \bp\ analog.
However, the lack of a significant amount of CS dust near the star argues against this scenario.
The process that is thought to give rise to the large numbers of star-grazing planetesimals in the \bp\ system, gravitational perturbation by an unseen giant planet \citep{Beust:1990}, would also be expected to cause dust-producing collisions between planetesimals. 
The time-variable, narrow absorption lines seen in spectra of HD~217782 seem to require a different explanation.
None of the four stars in our sample with claims for variable CS gas 
showed significant IR excess emission.
\citet{Redfield:2007b} found the same result using \si\ observations of 
three other stars with variable CS gas. 
The origin of time-variable, narrow absorption lines in spectra of main 
sequence early-type stars is a topic we plan to investigate further. 
%
%

\subsection{The Disk Fraction Among MS Shell Stars} \label{sub:diskfrac}

We surveyed 12 out of the 23 main sequence shell stars compiled in
\citet{Hauck:2000}.
Six of the 23 were already known to harbor debris disks or younger protoplanetary disks (MWC~480, \bp, HD~97048, HD~139614, HD~163296, \& HD~179218);
these stars were observed in other \si\ programs.
The remaining 5 stars were not observed in our program because they lie 
outside the LIB.
However, one of them (HD~144667) is an accreting Herbig~Ae star, which 
very likely has a protoplanetary disk \citep[e.g.][]{GarciaLopez:2006}.
Therefore, we consider that there were 7 known protoplanetary or debris disks in
the \citet{Hauck:2000} sample of main sequence shell stars prior to our \si\ survey. 

We found four additional disk candidates among the 12 main sequence shell stars from \citet{Hauck:2000} that we observed.
Therefore, the disk fraction among the \citet{Hauck:2000} main sequence
shell stars appears to be at least 11 out of 23 ($48\% \pm 14\%$).
We have no information about the presence of disks around four of the 23 stars; therefore, the disk fraction in this set of stars could be as high as $65\%$.

At this time, we are not able to properly compare the disk fraction among the main sequence shell stars to the disk fraction in other sets of early-type stars.
The disk fraction is higher for young early-type stars than it is for early-type stars with ages $> 30$~Myr \citep{Su:2006}.
Therefore, when comparing the disk fraction in different sets of stars, 
the sets should have similar distributions of stellar ages.
We do not currently know the age distribution in the \citet{Hauck:2000} main sequence shell star set, although we plan to estimate the stellar ages using multiple techniques in a future paper. 

\section{Concluding Remarks} \label{sec:conc} 

We have found four debris disk candidates in our \si\ survey of 16 main 
sequence shell stars. 
These stars show IR excess emission characteristic of moderate amounts of cool CS dust.
We have begun follow-up observations of all of our shell stars, to more fully characterize their CS material.

Based on their SEDs, it seems likely that our candidates will be confirmed as debris disks.
If so, they are likely to be edge-on, which is the most advantageous orientation for coronagraphic imaging of a disk.
These disk candidates will likely double the number of debris disks with currently detectable CS gas.
With this sample, we may begin a program to determine why some debris disks 
have both gas and dust and how these different components are related.

As mentioned in the Introduction, the nature of the main sequence shell stars has been something of a puzzle.
Our \si\ study suggests that at least half of the stars in this class have protoplanetary or debris disks.
Surveys for shell stars using ground-based optical spectroscopy, followed by IR photometry to look for CS dust, may be an effective way of finding protoplanetary and debris disks.

\acknowledgments

This work is based on observations made with the Spitzer Space Telescope, 
which is operated by the Jet Propulsion Laboratory, California Institute of Technology under a contract with NASA. 
Support for this work was provided by NASA through an award issued by 
JPL/Caltech.
This publication makes use of data products from the Two~Micron All Sky Survey
(2MASS), which is a joint project of the University of Massachusetts and the Infrared Processing and Analysis Center/California Institute of Technology, funded by the National Aeronautics and Space Administration and the National Science Foundation.



{\it Facilities:} \facility{Spitzer}.


\end{document}